\documentclass[graybox]{svmult}

\usepackage{mathptmx}       
\usepackage{makeidx}         
\usepackage{graphicx}        
\usepackage{multicol}        
\usepackage[bottom]{footmisc}
\usepackage{amsmath}
\usepackage{amssymb}
\usepackage{natbib}
\usepackage{hyperref}
\usepackage{bm,xstring}
\bibpunct{(}{)}{,}{a}{}{,}

\def\vect#1%
    {\IfSubStr{ABCDEFGHIJKLMNOPQRSTUVWXYZabcdefghijklmnopqrstuvwxyz}{#1}
        {\mathbf{#1}}
        {\bm{#1}}}

\makeindex             

\renewcommand{\dfrac}[2]{\frac{\mathrm{d} #1}{\mathrm{d} #2}}
\newcommand{\ddfrac}[2]{\frac{\mathrm{d}^2 #1}{\mathrm{d} #2^2}}

\renewcommand{\l}{\ell}
\renewcommand{\L}{\ell(\ell+1)}

\newcommand{\Kavg}{\mathcal{K}_{\mathrm{avg}}}
\newcommand{\Kcross}{\mathcal{K}_{\mathrm{cross}}}
\newcommand{\Kpavg}{\mathcal{K}'_{\mathrm{avg}}}
\newcommand{\Kpcross}{\mathcal{K}'_{\mathrm{cross}}}

\newcommand{\grad}{\vect{\nabla}}

\renewcommand{\div}{\vect{\nabla} \cdot}

\newcommand{\er}{\vect{e}_r}

\newcommand{\Ylm}{Y^{\ell}_m}

\begin{document}

\title*{Stellar Inversion Techniques}
\author{Daniel R.~Reese}

\institute{Daniel R.~Reese \at LESIA, Observatoire de Paris, PSL Research University, CNRS, Sorbonne Universit\'{e}s, UPMC Univ.~Paris 06, Univ.~Paris Diderot, Sorbonne Paris Cit\'{e}e, 92195 Meudon, France, \\ \email{daniel.reese@obspm.fr}}

\maketitle

\abstract{Stellar seismic inversions have proved to be a powerful technique
for probing the internal structure of stars, and paving the way for a better
understanding of the underlying physics by revealing some of the shortcomings
in current stellar models.  In this lecture, we provide an introduction
to this topic by explaining kernel-based inversion techniques.  Specifically,
we explain how various kernels are obtained from the pulsation equations,
and describe inversion techniques such as the Regularised Least-Squares (RLS)
and Optimally Localised Averages (OLA) methods.}

\section{Introduction}

Many of the problems which intervene in physics can be described in terms of
forward and inverse problems.  Generally speaking, a forward problem focuses on
predicting the effects which result from a set of physical causes, such as
deducing the gravitational field of an object from its distribution of matter. 
In an inverse problem, one typically tries to deduce the physical causes which
led to a given set of results or effects (which are typically observations). 
Hence, trying to deduce the distribution of matter from the gravitational field
of an object is an inverse problem.

The field of asteroseismology, i.e., the study of stellar pulsations, also fits
this description.  Trying to predict stellar pulsation frequencies for a given
stellar model constitutes a forward problem.  Likewise, trying to deduce the
stellar structure which led to a given set of pulsation frequencies is an
inverse problem.  This inverse problem turns out to be quite difficult because,
in general, the relation between stellar structure and oscillation frequencies
is non-linear.  Nonetheless, given the wealth of
information on the internal structure of stars provided by pulsation
frequencies, a variety of approaches have been devised to tackle this problem,
as expressed by \citet{Gough1985} in the context of helioseismology:
\begin{quotation}
Inversions can conveniently be divided into three categories.  The simplest
consists of the execution of the forward problem using solar models with a few
adjustable parameters, and the calibration of those parameters by fitting theory
to observation.  The second is the use of analytical methods. [...]  Thirdly,
there are the formal inversion techniques borrowed from geophysics that have
been used on real and artificial solar data.
\end{quotation}

The first category of inversions is usually named ``forward modelling'' (not to
be confused with the ``forward problem'') and corresponds to searching for an
optimal model in a restricted parameter space.  It typically includes methods
such as grid searches \citep[e.g.,][]{SilvaAguirre2015}, MCMC methods
\citep[e.g.,][]{Bazot2012}, or genetic algorithms
\citep{Metcalfe2003,Charpinet2005}.  The advantages of this approach is its
obvious simplicity, and the fact that it produces physically coherent models. 
However, the parameter space is restricted and does not allow for
hitherto unknown physical ingredients not included in the stellar models. 
Furthermore, such methods can be costly, especially if models are calculated
on-the-fly.  The second approach includes methods such as asymptotic methods or
glitch fitting.  These methods can provide a great deal of physical insight into
stellar physics but are beyond the scope of the present lecture.  Finally,
formal inversion techniques typically consist in adjusting the structure of a
reference stellar model so as to match a set of observed frequencies.  The
advantage of this approach is that it can potentially extract more information
from the pulsation frequencies, and is therefore open to new physics.  However,
this method may lead to models which are not physically coherent, and can be
more difficult to implement.  These approaches are in fact complementary. 
Indeed, the forward approach typically provides a reference model, which can
then be further refined via formal inversion techniques.

The present lecture focuses on the third category, i.e., formal inversion
techniques.  However, before tackling inversions, it is necessary to spend a bit
of time on the forward problem in order to bring out some of the properties
which apply in the context of inverse problems.  This will be the subject of the
next section.  Then stellar inversion techniques will be described in
Sect.~\ref{sect:inversions}.  A short conclusion including a list of relevant
references and available inversion codes will follow.

\section{The forward problem}

\subsection{Adiabatic pulsation equations}

Stellar pulsations, the periodic motion of
gas or plasma within a star, are described by the Lagrangian displacement and
the Eulerian perturbations to density, pressure, and gravitational potential,
denoted $\vect{\xi}$, $\rho'$, $p'$ and $\phi'$, respectively.  When applying
the adiabatic approximation (i.e., when neglecting heat transfers during the
periodic motions), these quantities are determined by Euler's equation, the
continuity equation, and the adiabatic relation, which express the conservation
of momentum, mass, and energy, as well as Poisson's equation.  Through various
analytical manipulations, and the use of Green's function and suitable boundary
conditions for Poisson's equation, it is possible to express $\rho'$, $p'$ and
$\phi'$ as a function of $\vect{\xi}$ alone. When inserted into Euler's
equation, this leads to the following schematic equation:
\begin{equation}
   \omega^2 \vect{\xi} = \mathcal{F}(\vect{\xi}) \, ,
   \label{eq:schematic}
\end{equation}
where $\mathcal{F}$ is an integro-differential operator, and where we have
assumed a time dependence\footnote{If one assumes that modes are proportional to
$\exp({\rm i}m\varphi)$, $\varphi$ being the longitude, such a time dependence will
lead to $m > 0$ modes being prograde, where $m$ is the azimuthal order.  If one
uses, instead, a time dependence of the form $\exp({\rm i}\omega t)$, then $m > 0$
modes will be retrograde.} of the form $\exp(-{\rm i}\omega t)$. 
Equation~(\ref{eq:schematic}), along with appropriate boundary conditions, is an
eigenvalue problem, the solutions of which are known as ``eigensolutions''. 
Specifically, $\omega^2$ is an eigenvalue and corresponds to the square of the
pulsation frequency, whereas $\vect{\xi}$ is the eigenmode or eigenfunction, and
specifies the geometric characteristics of the stellar pulsation.

The forward problem in this case, then corresponds to finding the above
eigensolutions for a given stellar structure, i.e., for a given $\mathcal{F}$
operator.  The inverse problem corresponds to finding the stellar structure (and
hence $\mathcal{F}$) from a set of pulsation frequencies and some sort of mode
identification, i.e., a partial characterisation of the structure of the
pulsation modes.  In the case of solar-like oscillators, a mode identification
typically includes the harmonic degrees $\l$ of the pulsations, and possibly the
radial orders $n$ (this is usually obtained from comparisons with models) and
azimuthal orders $m$ (only if frequency multiplets, typically caused by stellar
rotation, can be resolved).  Given that the forward problem is non-linear, the
inverse problem will also be non-linear.  However, to make the problem more
tractable, one typically linearises it.  Linearising Eq.~(\ref{eq:schematic})
leads to the following equation\footnote{Throughout these lectures, the $\delta$
notation will be used to indicate a modification of the equilibrium stellar
structure and associated pulsations.}:
\begin{equation}
   (\delta \omega^2) \vect{\xi} + \omega^2 (\delta \vect{\xi})
  = \delta \mathcal{F}(\vect{\xi}) + \mathcal{F}(\delta \vect{\xi}) \, .
\label{eq:linear_one}
\end{equation}
This equation simply expresses how a small modification to the stellar structure
leads to small modifications\footnote{It is very important to note that here,
the ``$\delta$'' symbol is not a Lagrangian perturbation, but rather a
modification of the model and its pulsations.} of the pulsation modes, in
particular frequency differences $\delta \omega$.  Hence, in order to solve the
inverse problem, one needs to find a reference stellar model (typically using
some form of forward modelling) which is sufficiently close to the true stellar
model so that the linear approximation applies, and \textit{invert} the
frequency differences, in order to find how to correct the stellar model so that
it more closely matches the actual star.  However, Eq.~(\ref{eq:linear_one}) is
not straightforward to use as it contains terms with $\delta \vect{\xi}$, the
perturbation of the eigenmode.  The next section shows how to remove these terms
by exploiting an important property of the adiabatic pulsation equations, namely
their symmetry.

\subsection{Symmetry of the adiabatic pulsation equations}

Before explaining in what sense the pulsation equations are symmetric, it is
necessary to introduce the following dot product:
\begin{equation}
\left<\vect{\eta},\vect{\xi}\right> = \int_V \rho_0 \vect{\eta}^* \cdot \vect{\xi} \mathrm{d}V \, ,
\label{eq:dot_product}
\end{equation}
where $\vect{\eta}^*$ is the complex conjugate of $\vect{\eta}$, and $V$ the
stellar volume.  We note that this is a complex dot product, hence:
$\left<\vect{\eta},\vect{\xi}\right> = \left<\vect{\xi},\vect{\eta}\right>^*$.

The adiabatic pulsation equations are symmetric with respect to the above dot
product:
\begin{equation}
    \left< \vect{\eta}, \mathcal{F}(\vect{\xi}) \right>
  = \left< \mathcal{F}(\vect{\eta}), \vect{\xi} \right> \, ,
\end{equation}
where $\vect{\xi}$ and $\vect{\eta}$ are any displacement fields, which need not
necessarily be eigenfunctions at this point.  In order to prove this symmetry,
we start by introducing the associated pressure and gravitational potential
perturbations as deduced from the relevant equations: $(\vect{\xi},p',\phi')$
and $(\vect{\eta},\pi',\psi')$.   We then calculate the dot product between
$\vect{\eta}$ and Euler's equation (applied to $\vect{\xi}$). After various
manipulations (integration by parts etc.), this leads to the following formula
\citep[e.g.,][]{Unno1989}:
\begin{eqnarray}
\left< \vect{\eta}, \mathcal{F}(\vect{\xi}) \right>
  &=& \int_V \frac{(\pi')^* p'}{\rho_0 c_0^2} \mathrm{d}V 
   + \int_V \rho_0 N_0^2 (\vect{\eta}^*\cdot \er) (\vect{\xi} \cdot \er) \mathrm{d}V \nonumber \\
  &+& \int_S \rho_0 g_0 (\vect{\eta}^*\cdot \er) (\vect{\xi} \cdot \er) \mathrm{d}S
   - \frac{1}{4\pi G} \int_{V_{\infty}} \grad (\psi')^* \cdot \grad \phi' \mathrm{d}V \, ,
\end{eqnarray}
where $V$ is the star's volume, $S$ its surface, $V_{\infty}$ infinite space,
$\er$ the unit vector in the radial direction,
and $N_0^2$ the square of the Brunt--V{\"a}is{\"a}l{\"a} frequency.  In deriving
the surface term, we assumed, as a boundary condition, that the Lagrangian
pressure perturbation vanishes at the surface.  Appendix C of
\citet{Reese2006phd} explains how to obtain the last term (integrated over
$V_{\infty}$).  It is very clear from this explicit formulation that the
pulsation equations are symmetric.  More general forms of this equation have
been derived, for instance, in the case of differentially rotating physical
bodies \citep{Lynden-Bell1967}.

This symmetry leads to a number of consequences.  Firstly, the eigenvalues,
$\omega^2$, are real (meaning that the $\omega$ are either real or purely
imaginary).  Secondly, the eigenfunctions of distinct eigenvalues are orthogonal
with respect to the above dot product.  The third consequence is known as the
``variational principle''.  According to this principle, the variational
frequency, defined by $ \omega^2_{\mathrm{var}} = \left< \vect{\xi},
\mathcal{F}(\vect{\xi})\right> / \left< \vect{\xi}, \vect{\xi}\right>$, differs
from the true eigenfrequency by an amount which is of second order or higher in
terms of the error on the eigenfunction, i.e., $\omega^2-\omega_{\mathrm{var}}^2
= \mathcal{O}\left(\|\Delta\vect{\xi}\|^2\right)$. This is useful as
$\omega^2_{\mathrm{var}}$ provides an independent and potentially more accurate
estimate of the eigenvalue than the numerical value and is therefore used 
as an accuracy test in various pulsation codes such as \textsc{adipls}
\citep{Christensen-Dalsgaard2008}.

\subsection{Kernels}
\label{sect:kernels}

We now return to our original problem, i.e., calculating the frequency variation
caused by a small modification of the stellar structure. Taking the dot product
between Eq.~(\ref{eq:linear_one}) and $\vect{\xi}$, and grouping terms
with $\delta\vect{\xi}$ yields:
\begin{equation}
     \delta \omega^2 \left<\vect{\xi},\vect{\xi}\right>
    - \left<\vect{\xi},\delta \mathcal{F}(\vect{\xi})\right> 
   = \left<-\omega^2 \vect{\xi}+\mathcal{F}(\vect{\xi}),\delta \vect{\xi}\right> \, ,
\end{equation}
where we have made use of the symmetry of $\mathcal{F}$. The right-hand side
vanishes because $\vect{\xi}$ is an eigenmode, and $\omega^2$ the corresponding
eigenvalue.  Isolating $\delta \omega^2$ then yields:
\begin{equation}
  \delta \omega^2 = 2 \omega \delta \omega =
  \frac{\left<\vect{\xi},\delta\mathcal{F}(\vect{\xi})\right>}
       {\left<\vect{\xi},\vect{\xi}\right>} \, .
\end{equation}
This last form is extremely useful because it relates modifications of the
pulsation frequency directly to changes in the stellar model, \textit{without
needing} $\delta \vect{\xi}$.

The next obvious question is what types of perturbations can we expect in stars?
A first type of perturbation, which in fact is ubiquitous, is stellar rotation. 
One can distinguish the 1D case, where the rotation profile, $\Omega$, only
depends on the radial coordinate $r$ (also known as ``shellular'' rotation) from
the 2D case where it depends on $r$ and $\theta$, the colatitude.  A second type
of perturbation is modifications to the stellar structure, as defined, for
instance, by the $\rho_0$, $\Gamma_{1,0}$, $c_0^2$ etc., profiles.  So far,
structural modifications have only been envisaged in a 1D setting.

\begin{figure}[t]
\centering
\includegraphics[width=0.48\textwidth]{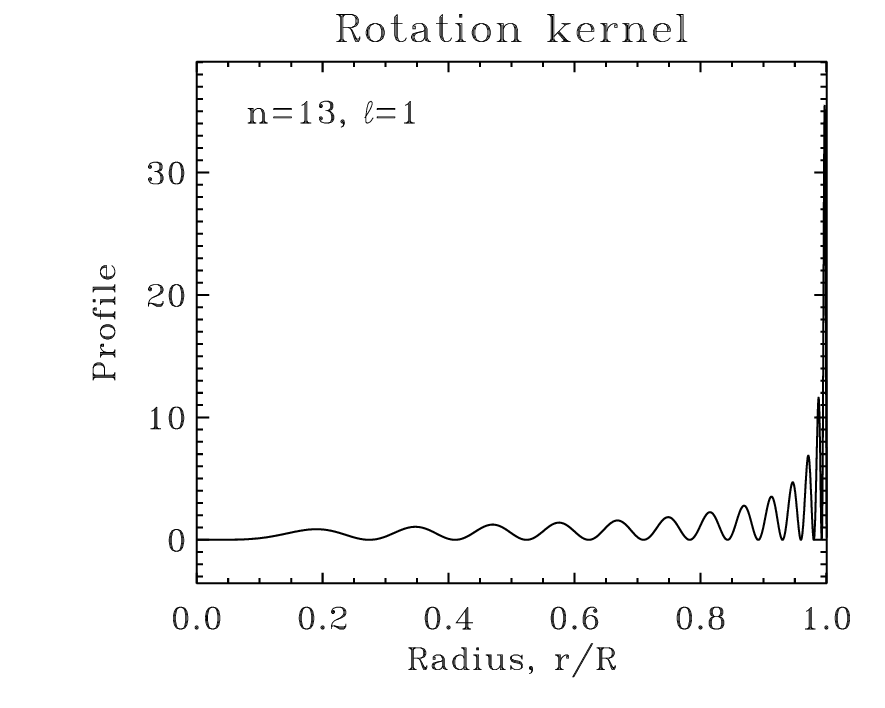}
\includegraphics[width=0.48\textwidth]{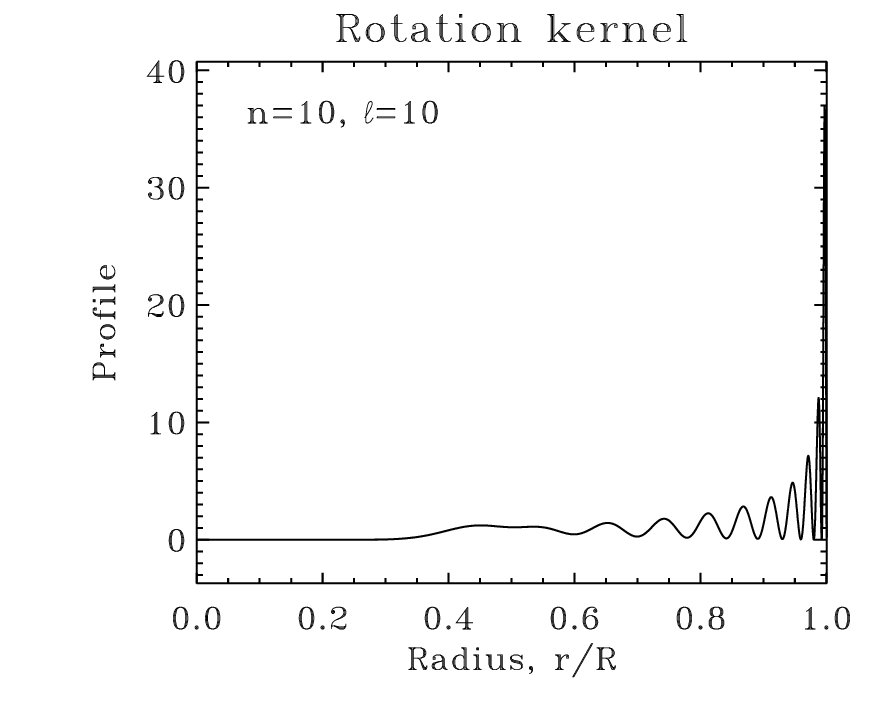}
\caption{Examples of 1D rotation kernels. \label{fig:rotation_kernels_1D}}.
\end{figure}

Rotation leads to two inertial accelerations: the centrifugal and the Coriolis
acceleration.  The former distorts the shape of the star but is a
second order effect, so will be neglected.  The latter intervenes in
the oscillatory motions and leads to first order effects on the frequencies.  To
first order, Euler's equation takes on the form:
\begin{equation}
   \omega^2 \vect{\xi} =  2 \omega m \Omega \vect{\xi}
                        - 2i\omega \vect{\Omega} \times \vect{\xi}
     + \frac{\grad p'}{\rho_0} - \frac{\rho' \vect{g}_0}{\rho_0} + \grad \phi' \, ,
\end{equation}
where $\Omega$ is the rotation profile, and $\vect{\Omega} = \Omega
\vect{e}_z$.  From this we deduce:
\begin{equation}
  \delta \mathcal{F}(\vect{\xi}) =  2 \omega m \Omega \vect{\xi}
                                 - 2i\omega \vect{\Omega} \times \vect{\xi} \, .
\end{equation}
In the 1D case, the frequency shift is given by
\begin{equation}
  \delta \omega_{n,\,\l,\,m} = \omega_{n,\,\l,\,m} - \omega_{n,\,\l,\,0}
                             = m\int_0^R K_{\Omega}^{n\l}(r) \Omega(r) \mathrm{d}R \, ,
\label{eq:rotation_splitting_1D}
\end{equation}
where
\begin{equation}
K_{\Omega}^{n\l} = \frac{\rho_0 r^2\left(\xi_r^2+\L\xi_h^2-2\xi_r\xi_h-\xi_h^2\right)}
                          {\int_{0}^{R} \rho_0(r) \left(\xi_r^2 + \l(\l+1) \xi_h^2\right) r^2 \mathrm{d}r} 
\end{equation}
and where $\xi_r$ and $\xi_h$ are the radial and horizontal components of the
Lagrangian displacement, respectively.
$K_{\Omega}^{n\l}$ is known as the ``rotation kernel''.  As can be seen
from this expression, frequencies with the same $(n,\,\l)$ values are
uniformly split as a function of $m$ thanks to rotation. 
Figure~\ref{fig:rotation_kernels_1D} shows some examples of 1D rotation kernels.

If $\Omega$ is constant, then Eq.~(\ref{eq:rotation_splitting_1D}) simplifies to
$ \delta \omega = m(1 - \mathcal{C})\Omega $, where
\begin{equation}
    \mathcal{C} = \frac{\int_{0}^{R}\rho_0 \left(2\xi_r\xi_h+\xi_h^2\right) r^2\mathrm{d}r}
                  {\int_{0}^{R} \rho_0(r) \left(\xi_r^2 + \l(\l+1) \xi_h^2\right) r^2 \mathrm{d}r} \, .
\end{equation}
$\mathcal{C}$ is known as the Ledoux constant and represents the effects of the
Coriolis force \citep[see][]{Ledoux1951}.

In the 2D case, the rotational splitting is given by
\begin{equation}
\delta \omega_{n,\,\l,\,m} 
= \int_0^R \int_0^\pi \mathcal{K}_{n,\,\l,\,m}(r,\theta) \Omega(r,\theta)r
\mathrm{d}r\mathrm{d}\theta \, ,
\end{equation}
where $\mathcal{K}_{n,\,\l\,\,m}(r,\theta)$ is the 2D rotation kernel
\citep[expressions for such kernels may be found in][]{Schou1994}.
This time, the splitting as a function of $m$ may be non-uniform.

\begin{figure}[t]
  \centering
  \includegraphics[width=0.48\textwidth]{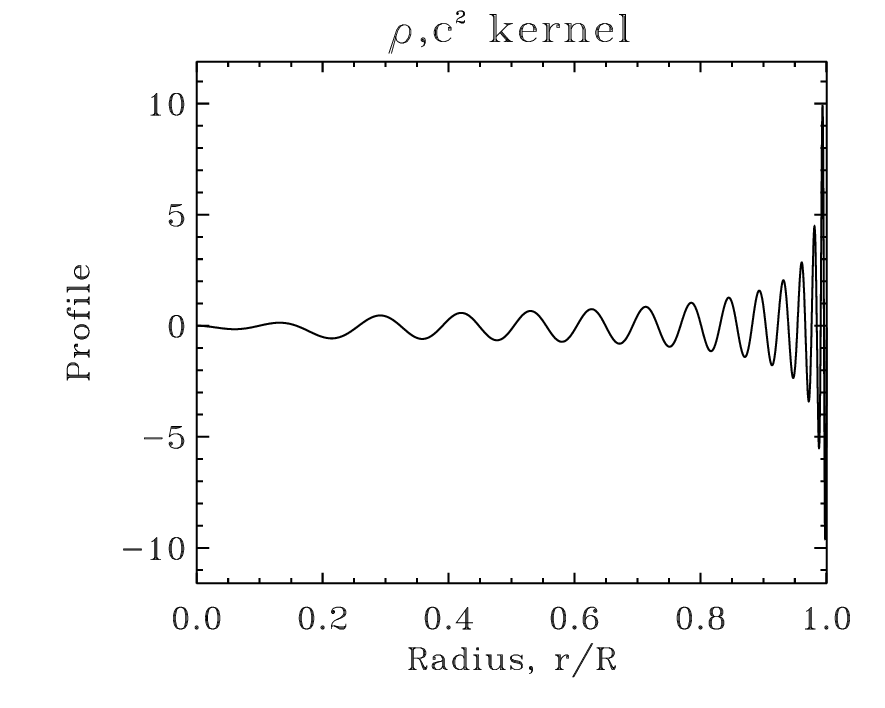} 
  \includegraphics[width=0.48\textwidth]{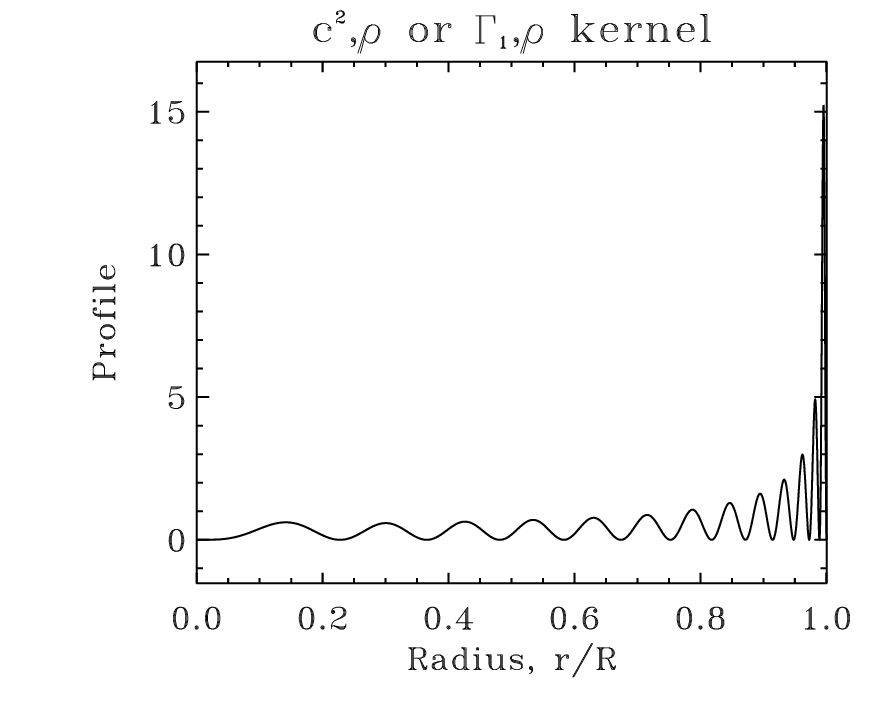} 
  \includegraphics[width=0.48\textwidth]{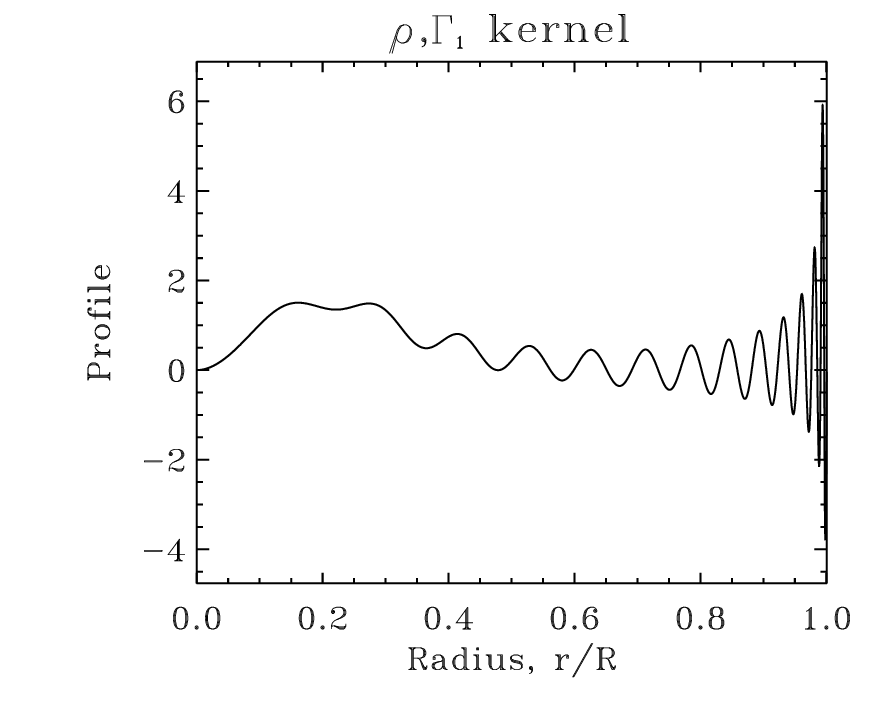}
  \caption{Kernels for the $(n,\l)=(13,1)$ pulsation mode, 
           for the structural pairs $(\rho_0,c_0^2)$
           and $(\rho_0,\Gamma_{1,0})$. \label{fig:structural_kernels}}
\end{figure}

The acoustic structure of stars is typically determined by two variables, e.g., 
$(\rho_0, \Gamma_{1,0})$ and possibly some surface quantities such as the
surface pressure.  Accordingly, when modifying the structure of a star, the
modifications to two structural quantities need to be specified, e.g., 
$(\delta\rho_0, \delta\Gamma_{1,0})$ \citep[although in some cases, 3 functions need
to be specified, e.g.,][]{Buldgen2017}.  As was the case for rotation, it
is possible to relate changes in frequency to structural modifications of stars
using kernels.  The easiest structural kernels to derive are those for the
variables $(\rho, c^2)$.  After a (very) lengthy derivation, one can show
that
\begin{equation}
  \frac{\delta \omega}{\omega} = \int_{0}^R \left[K_{c^2,\rho}(r)\frac{\delta c^2_0(r)}{c^2_0(r)}
                               + K_{\rho,c^2}(r)\frac{\delta\rho_0(r)}{\rho_0(r)}\right] \mathrm{d}r \, ,
\label{eq:rho_c2_kernels}
\end{equation}
where:
\begin{eqnarray}
  K_{c^2,\rho} &=& \frac{\rho_0 c_0^2 \chi^2 r^2}{2I\omega^2} \, , \\
  K_{\rho,c^2} &=& \frac{\rho_0 r^2}{2I\omega^2} \left\{ c_0^2 \chi^2 
                 -\omega^2 \left(\xi_r^2+\L\xi_h^2\right)
                 -4\pi G \int_{s=r}^R \left(2\rho_0\xi_r\chi+\dfrac{\rho_0}{s}\xi_r^2\right)\mathrm{d}s \right. \nonumber \\
               & & \left.  -2g_0\xi_r\chi +2g_0\xi_r\dfrac{\xi_r}{r}+4\pi G\rho_0\xi_r^2
                 +2\left(\xi_r \dfrac{\phi'}{r} + \frac{\L \xi_h \phi'}{r}\right) \right\} \, , \\
  I          &=& \int_0^R \rho_0 \left(\xi_r^2 + \l(\l+1)\xi_h^2\right) r^2 \mathrm{d}r \; , \;\;
  \chi       = \frac{\div \vect{\xi}}{\Ylm} = \dfrac{\xi_r}{r} + \frac{2\xi_r}{r} - \frac{\L\xi_h}{r} \, . \nonumber
\end{eqnarray}
Figure~\ref{fig:structural_kernels} gives an example of $(\rho,c^2)$
kernels. We note that in deriving Eq.~(\ref{eq:rho_c2_kernels}), we neglected
various surface terms which result from integration by parts.  Also, the
modelling of surface layers in stars tends to be inaccurate.  Accordingly,
Eq.~(\ref{eq:rho_c2_kernels}) typically includes an extra ad hoc adjustable
surface term.

Besides these kernels, other structural kernels can also be obtained:
$(\rho,\Gamma_1)$, $(P,\Gamma_1)$, $(u \equiv \frac{P}{\rho},\Gamma_1)$,
$(g,\Gamma_1)$,  $(u,Y)$, $(A,\Gamma_1)$, $(N^2,c^2)$ etc.~
\citep[see][]{Masters1979, Gough1991, Elliott1996, Basu1997, Kosovichev1999,
Buldgen2017}. Some of these require using the equation of state and its
derivatives.  Figure~\ref{fig:structural_kernels} shows an example of
$(\rho,\Gamma_1)$ kernels.

\section{The inverse problem}
\label{sect:inversions}

As described at the beginning of this lecture, the seismic inverse problem
consists in deducing the stellar structure from a set of \textit{identified}
pulsation frequencies, i.e., with known quantum numbers.  Inverse methods have
proved to be a powerful way of solving such a problem.  These typically involve
correcting a reference stellar model so as to obtain a new model which
reproduces the pulsation frequencies more accurately.  Inverse methods come into
two broad categories, namely linear and non-linear methods.  The linear methods
are further subdivided into the Regularised Least-Squares (RLS) and Optimally
Localised Averages (OLA) methods.  For the non-linear inversions, there are
iterated versions of the RLS method, as well as a method which adjusts the
internal phases of the eigenmodes.  In what follows, we will focus on linear
inverse methods, beginning with rotation inversions, as these provide a good
starting point to illustrate the different methods.

\subsection{Rotation inversions}

The rotation inverse problem can be expressed by the following set of equations:
\begin{equation}
  S_{n_l,\l_l} = \frac{\omega_{n_l,\l_l,m_l} - \omega_{n_l,\l_l,0}}{m_l}
               = \int_0^R K_{\Omega}^{n_l,\l_l}(r) \Omega(r)\mathrm{d}r
               + \varepsilon_{n_l,\l_l}\,, \;\; 1 \leq l \leq L\,,
\label{eq:rota_inverse_problem}
\end{equation}
where the $S_{n_l,\l_l}$ are the ``rotational splittings'' (i.e., the
observations), $\Omega(r)$ the unknown rotation profile, and
$\varepsilon_{n_l,\l_l}$ the errors on the splittings, characterised by a
standard deviation of $\sigma_{n_l,\l_l} = \left< \varepsilon_{n_l,\l_l}
\right>$.  In what follows, we will use the index ``$l$'' as shorthand for
$(n_l,\l_l)$.

The goal of the inverse problem is to recover $\Omega(r)$ from the set of
available rotational splittings.  At first, this problem looks impossible. 
Indeed, the unknown is a function, whereas there is a finite number of
observational constraints. Furthermore, the problem is ill-conditioned, i.e., it
is highly sensitive to noise.  In order to address these issues, it is necessary
to inject \textit{a priori} assumptions when solving the inverse problem. 
Accordingly, we should always bear in mind these assumptions when looking at and
interpreting the results.

\subsubsection{Regularised Least Squares (RLS)}

A first approach to tackling this problem involves decomposing the rotation
profile over a set of basis functions:
\begin{equation}
    \Omega_{\mathrm{inv}}(r) = \sum_k a_k f_k(r) \, ,
\end{equation}
where the $a_k$ are unknown coefficients, and the $f_k$ basis functions. In
general, the number of unknown coefficients should be equal to or less than the
number of observed splittings. Typical choices for the $f_k$ include b-spline
functions of various degrees. For instance, zeroth degree b-splines produce
step-wise functions, whereas cubic splines produce functions with a continuous
second derivative (which can be useful for regularisation terms).

\begin{figure}[t]
\centering
  \includegraphics[width=0.8\textwidth]{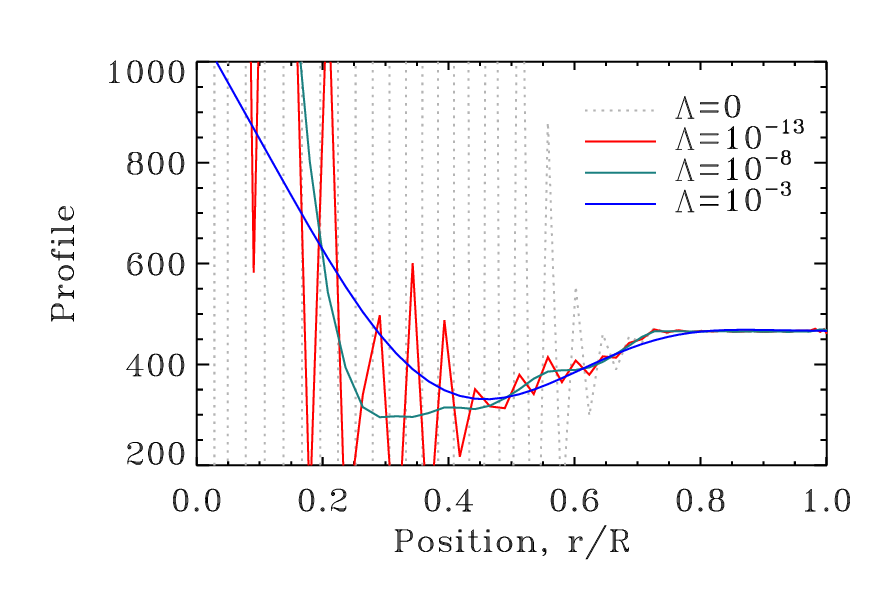} 
  \includegraphics[width=0.8\textwidth]{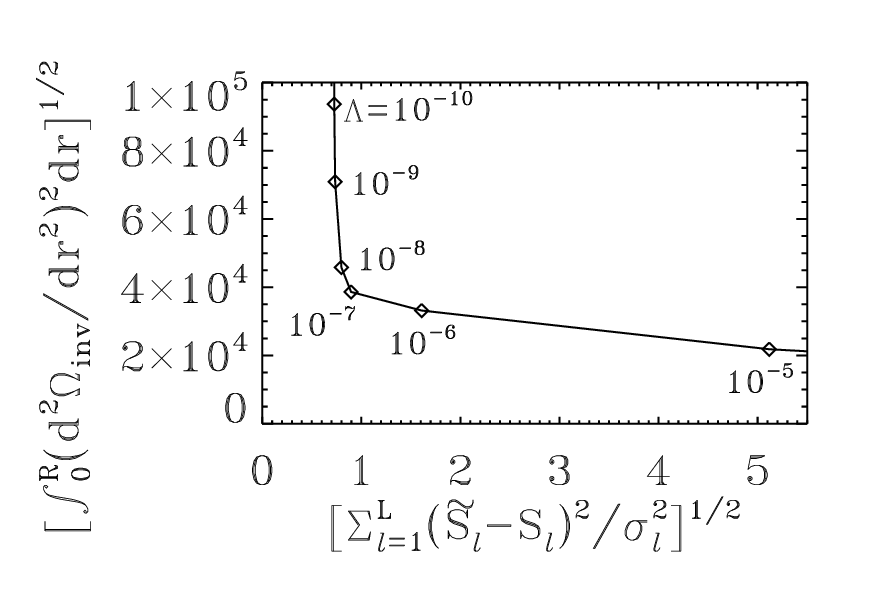}
  \caption{\textit{Top:} Inverted rotation profiles based on the RLS method for
  different values of the regularisation parameter. \textit{Bottom:} L-curve which
  shows the two components of the RLS cost function as a function of $\Lambda$.
  \label{fig:RLS_rota}}
\end{figure}

When substituted into Eq.~(\ref{eq:rota_inverse_problem}), this leads to the
following theoretical rotational splittings, $\tilde{S}_{l}$, for the above
rotation profile:
\begin{equation}
  \tilde{S}_{l} = \int_0^R K_{\Omega}^{l}(r) \Omega_{\mathrm{inv}}(r)\mathrm{d}r \, .
\end{equation}
An obvious way of choosing the $a_k$ is by minimising (typically in a
least-squares sense) the distance between the observed splittings, $S_l$, and
the theoretical ones.  However, a naive application of such a procedure leads to
poor results as illustrated by the dotted grey curve in the top panel of
Fig.~\ref{fig:RLS_rota}.  Indeed, the problem is ill-conditioned, and any
errors in the observations will be strongly amplified.

A standard remedy to this problem is to include a supplementary regularisation
term to obtain a smooth solution when carrying out the minimisation, hence the
name ``Regularised Least-Squares'' (RLS) method.  This leads to the following
typical cost function:
\begin{equation}
     J(a_k) = \sum_{l=1}^{L} \frac{\left(S_{l} - \tilde{S}_{l}\right)^2}{\sigma_l^2}
                + \Lambda \left< \frac{1}{\sigma^2} \right>
                  \int_0^R \left(\ddfrac{\Omega_{\mathrm{inv}}}{r}\right)^2 \mathrm{d}r \, ,
\end{equation}
where $\left< \frac{1}{\sigma^2} \right> = \frac{1}{L} \sum_{l=1}^{L} \frac{1}
{\sigma_l^2}$, and $\Lambda$ is a regularisation parameter which can be
adjusted.  The cost function is minimised by numerically finding the $a_k$
coefficients for which the gradient of $J$ is zero.


Figure~\ref{fig:RLS_rota} shows various solutions obtained for the rotation
inverse problem based on a set of rotational splittings from
\citet{Christensen-Dalsgaard1990}.  As can be seen in the top panel, larger
values of $\Lambda$ lead to solutions that are smoother.  The bottom panel shows
that such solutions are a worse fit to the $S_l$.  Hence, there is a trade-off
between obtaining smooth solutions and fitting the data.  The best solutions are
obtained for intermediate values of $\Lambda$ as can be seen by comparing the
solutions in Fig.~\ref{fig:RLS_rota} to the true solution given in figs.~3
and 11 of \citet{Christensen-Dalsgaard1990}.

\subsubsection{Various error measurements}

It is also possible to calculate error bars around the inverted solution.  To
demonstrate this, we start with a given grid point, $r_0$.  The relationship
between $\Omega_{\mathrm{inv}}(r_0)$ and the $a_k$ coefficients is linear. 
Likewise, the relationship between the $a_k$ and the $S_l$ is also linear. 
Hence the relationship between $\Omega_{\mathrm{inv}}(r_0)$  and the $S_l$
is linear and can be expressed as follows:
\begin{equation}
     \Omega_{\mathrm{inv}}(r_0) = \sum_l c_l(r_0) S_l \, .
     \label{eq:inversion_coefficients_RLS}
\end{equation}
Assuming the errors on the splittings are uncorrelated, the $1\sigma$ error bar
on the inverted value of the rotation rate will simply be
\begin{equation}
    \sigma_{\Omega(r_0)} = \sqrt{ \sum_l \left(c_l(r_0) \sigma_l\right)^2} \, .
\end{equation}
In the specific case where the errors are uniform, the error is amplified by the
quantity $\sqrt{\sum_l \left(c_l(r_0)\right)^2}$ which is known as the ``error
magnification''. It is important to bear in mind that these error bars only take
into account how the observational errors propagate through the inversion.  They
do not actually measure the quality of the inversion, which could, for example,
be poor due to over-regularisation.

\begin{figure}[t]
\centering
  \includegraphics[width=\textwidth]{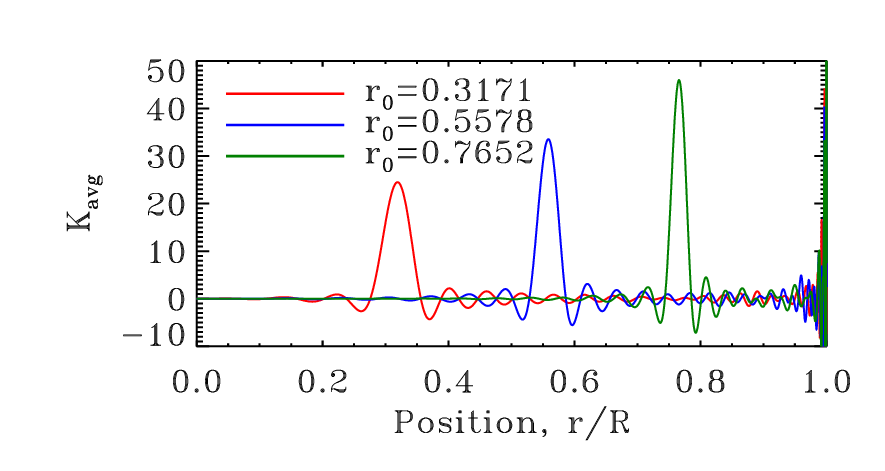}
  \caption{Averaging kernels for the RLS method at various positions.
           \label{fig:averaging_kernels_RLS}}
\end{figure}

In order to evaluate the quality of the inversion at a given point, it is useful
to look at the ``averaging kernel''.  If we replace the $S_l$ in
Eq.~(\ref{eq:inversion_coefficients_RLS}) by the expressions given in
Eq.~(\ref{eq:rota_inverse_problem}), then it is possible to establish a
relationship between $\Omega(r)$ and $\Omega_{\mathrm{inv}}(r_0)$:
\begin{equation}
   \Omega_{\mathrm{inv}}(r_0)
         = \int_0^R \underbrace{\sum_l c_l(r_0) K_{\Omega}^l(r)}_{\Kavg(r_0,r)}
             \Omega(r)\mathrm{d}r + \sum_l c_l(r_0) \varepsilon_l \, .
\end{equation}
This expression shows that $\Omega_{\mathrm{inv}}(r_0)$ is in fact an average of
the true rotation profile $\Omega(r)$.  The corresponding weight function,
$\Kavg(r_0,r)$, is the averaging kernel.  Ideally, this
function should have a strong amplitude at $r_0$ and be close to zero elsewhere.
Figure~\ref{fig:averaging_kernels_RLS} shows a few examples of averaging kernels
for the RLS method.

\subsubsection{Optimally Localised Averages (OLA)}

The notion of averaging kernels naturally leads to the Optimally Localised
Averages (OLA) methods.  The basic idea in these methods is to optimise the
coefficients $c_l$ so as to obtain \textit{optimal} averaging kernels. Two
variants include the Multiplicative and the Subtractive OLA, abbreviated MOLA and
SOLA, respectively.

The MOLA method comes from \citet{Backus1968}.  In this method, the averaging
kernel is \textit{multiplied} by a penalty function that increases in amplitude
as you move away from the target position $r_0$.  Hence, the coefficients
$c_l(r_0)$ are obtained by minimising the following cost function:
\begin{equation}
J(c_l) = \! \underbrace{\int_0^R  \!\!\!\! P(r_0,r) \left[\Kavg(r_0,r) \right]^2 \! \mathrm{d}r}_{\mbox{fit data}}
      + \underbrace{\frac{\tan\theta}{\left<\sigma^2\right>} \sum_{l=1}^L \left(c_l \sigma_l\right)^2}_{\mbox{regularisation}}
      + \underbrace{\lambda \left\{1- \int_0^R \!\!\!\! \Kavg \right\}}_{\mbox{$\Kavg$ unimodular}} \, ,
\end{equation}
where $\left< \sigma^2 \right> = \frac{1}{L}\sum_{l=1}^{L} \sigma_l^2$, $\theta$
is a trade-off parameter between fitting data and reducing error (i.e., a
regularisation parameter), $P(r_0,r)$ the penalty function (usually
$12(r-r_0)^2$), and $\lambda$ a Lagrange multiplier used to ensure that the
averaging kernel is ``unimodular'', i.e., $\int_0^R \Kavg(r_0,r)\mathrm{d}r = 1$.
This last condition is important for ensuring that the inverted value,
$\Omega_{\mathrm{inv}}(r_0) = \sum_l c_l(r_0) S_l$ is a proper average of the
underlying rotation profile.

The SOLA method was first described in \citet{Pijpers1992}.  In this method, the
\textit{difference} between the averaging kernel and a suitable target function
is minimised.  Hence, the coefficients $c_l(r_0)$ are obtained by minimising the
following cost function:
\begin{equation}
    J(c_l) = \! \int_0^R \!\!\!\! \big[ \mathcal{T}(r_0,r) - \Kavg(r_0,r) \big]^2 \mathrm{d}r 
      + \frac{\tan\theta}{\left<\sigma^2\right>} \sum_{l=1}^L \left(c_l \sigma_l\right)^2
      + \lambda \left\{1- \int_0^R \!\!\!\! \Kavg \right\} \, ,
\end{equation}
where $\mathcal{T}(r_0,r)$ is the target function.  Ideally, $\mathcal{T}$
should be a Dirac function centred on $r_0$.  However, given the finite number
of rotation splittings and hence rotation kernels to work with, trying to
achieve such a target is impossible and would lead to poor numerical results. 
Generally, Gaussian or similar functions are used as targets:
\begin{equation}
    \mathcal{T}(r_0,r) = \frac{1}{A} \exp\left(-\frac{(r-r_0)^2}{2\Delta(r_0)^2}\right) \, ,
\end{equation}
where $A$ is a normalisation constant to ensure that $\int_0^R
\mathcal{T}(r_0,r) \mathrm{d}r = 1$ (it is not simply $1/\sqrt{2\pi}\Delta(r_0)$
since the integration interval is not from $-\infty$ to $\infty$), and
$\Delta(r_0)$ the width of the target function.  A good choice for
$\Delta(r_0)$ when dealing with acoustic modes is $\Delta(r_0) \propto
c_0(r_0)$ \citep[e.g.,][]{Thompson1993}.

\begin{figure}[t]
\centering
  \includegraphics[width=0.8\textwidth]{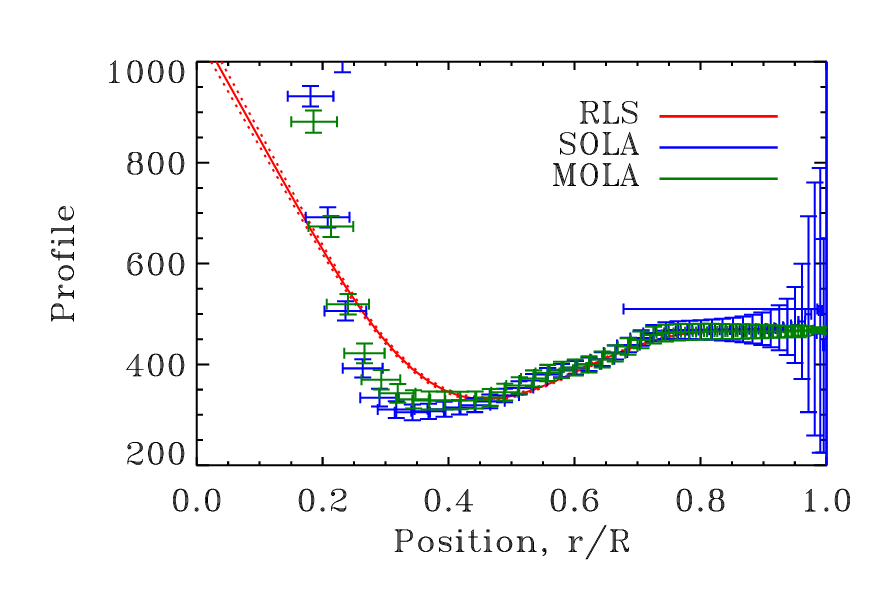} 
  \includegraphics[width=0.8\textwidth]{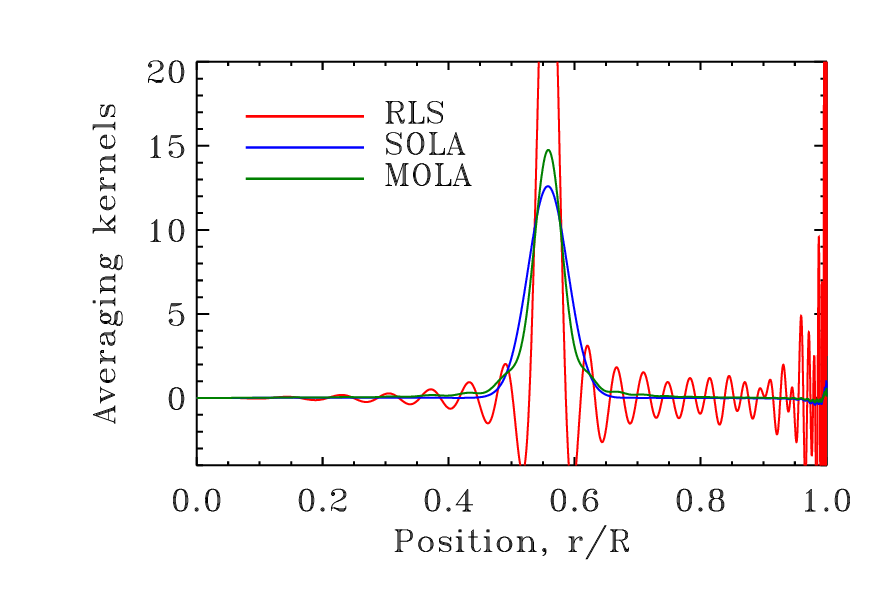} 
  \caption{\textit{Top:} Inversion results and error bars for the RLS, MOLA and SOLA methods. \textit{Bottom:} Averaging kernels at $r_0=0.5578\,R$ for these three methods.
           \label{fig:OLA_inversions}}
\end{figure}

Figure~\ref{fig:OLA_inversions} shows inversion results for the RLS, MOLA and
SOLA methods as well as some averaging kernels.  The advantages of the MOLA
method compared to the SOLA method is that it has fewer free parameters and
tends to produce slightly better results.  Conversely, the SOLA method has
a much smaller computational cost.  Indeed, minimising the SOLA cost function
for different values of $r_0$ leads to systems of equations where only the
right-hand side changes.  Accordingly, only one matrix inversion (or
factorisation) is needed for the entire inversion.

\subsubsection{Applications}

\begin{figure}[t]
\sidecaption
\includegraphics[width=0.55\textwidth]{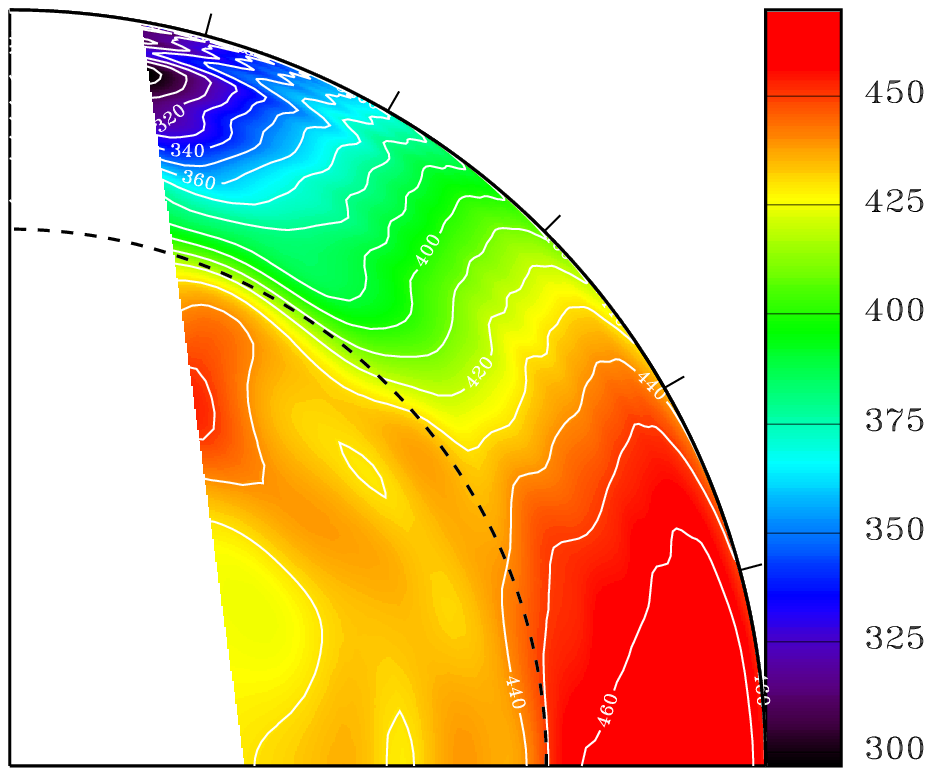}
\begin{minipage}{10mm}
\rotatebox{90}{nHz}
\vspace*{50mm}
\end{minipage}
\caption{2D solar rotation profile from \citet{Thompson2003} \citep[see also][]{Schou1998}
based on a SOLA inversion technique. Figure courtesy of M.~J.~Thompson and J.~Christensen-Dalsgaard.
\label{fig:solar_rotation}}
\vspace{-75pt}
\end{figure}

The first and most spectacular examples of rotation profile inversions are those
done for the Sun.  Indeed, the Sun's close proximity has enabled the detection
of countless rotational splittings going to high $\l$ values.  This, in turn,
has enabled 2D inversions of the solar rotation profile such as the one
shown in Fig.~\ref{fig:solar_rotation}, taken from \citet{Thompson2003}
\citep[see also][]{Schou1998}.  Such profiles were not in agreement with the
theoretical predictions at the time and have accordingly led to various
theoretical investigations and numerical simulations to gain a better understanding
of the Sun and its internal rotation \citep[e.g.,][]{Thompson2003, Brun2004}.

A more recent example of stellar rotation inversions are those in subgiants and
red giants \citep{Deheuvels2012, Deheuvels2014}.  These results as well as
results from ensemble asteroseismology have shown that although the core of these
stars rotate much faster than the envelope, the difference in rotation speeds is
orders of magnitude smaller than what is expected theoretically
\citep{Eggenberger2012, Ceillier2013, Marques2013}.  It is still an open
question what transport mechanisms are involved in these stars and could solve
this discrepancy.

\subsection{Structural inversions}

We now turn our attention to structural inversions.  In contrast to rotation
inversions, there are \textit{two} functions to invert simultaneously.  As was
derived in Sect.~\ref{sect:kernels}, the linearised relationship between
modifications of the stellar structure and shifts in the frequency can be
expressed as follows:
\begin{equation}
  \underbrace{\frac{\delta\omega_l}{\omega_l}}_{\mbox{obs.}} = 
    \int_0^R \underbrace{K_{a,b}^l (r)}_{\mbox{known}} \underbrace{\frac{\delta a}{a}}_{\mbox{unknown}} \mathrm{d}r
  + \int_0^R \underbrace{K_{b,a}^l (r)}_{\mbox{known}} \underbrace{\frac{\delta b}{b}}_{\mbox{unknown}} \mathrm{d}r
  + \frac{F_{\mathrm{surf.}}(\omega_l)}{E_l} + \varepsilon_l \, ,
\end{equation}
where we have added an ad hoc surface correction term (i.e., the
term with $F_{\mathrm{surf.}}$) as well as the observational error,
$\varepsilon_l$.  The variables $(a,b)$ represent two structural profiles (e.g.,
$(\rho,\Gamma_1)$).  The structural inverse problem then consists in deducing
the profiles $\delta a/a$ and $\delta b/b$ from the frequency shifts
$\delta \omega_l/\omega_l$.  The fact that there are two functions to invert leads to
modifications of the RLS and OLA methods, as well as the introduction of
``cross-term kernels'', $\Kcross$.

\subsubsection{Regularised Least Squares (RLS)}

In the regularised least squares method, both functions ($\delta a/a$ and
$\delta b/b$) are discretised over a set of basis functions, and the unknown
coefficients are obtained by minimising a cost function of the form:
\begin{eqnarray}
     J\left(\frac{\delta a}{a}, \frac{\delta b}{b}\right)
               &=& \sum_{l} \frac{1}{\sigma_l^2}\left(\frac{\delta\omega_l}{\omega_l} 
                - \int_0^R K_{a,b}^l \frac{\delta a}{a} \mathrm{d}r
                - \int_0^R K_{b,a}^l \frac{\delta b}{b} \mathrm{d}r\right)^2
                \nonumber \\
               & &+ \Lambda \left< \frac{1}{\sigma^2} \right>
                   \int_0^R \left[\left(\ddfrac{}{r}\frac{\delta a}{a}\right)^2 +
                                  \left(\ddfrac{}{r}\frac{\delta b}{b}\right)^2\right]\mathrm{d}r \, .
\end{eqnarray}
Additional terms may be included to model surface effects.

In much the same way as for rotation inversions, the inverted functions are
related in a linear way to the observables $\left(\delta
\omega/\omega\right)_{l}$:
\begin{equation}
    \left(\frac{\delta a}{a}\right)_{\mathrm{inv}} = \sum_l c_l(r_0) \left(\frac{\delta\omega}{\omega}\right)_l \; , \;\;
    \left(\frac{\delta b}{b}\right)_{\mathrm{inv}} = \sum_l c'_l(r_0) \left(\frac{\delta\omega}{\omega}\right)_l \, .
\end{equation}
These inversion coefficients can then be used to define the averaging
and cross-term kernels:
\begin{eqnarray}
\Kavg(r_0,r)   &=& \sum_{l=1}^{L} c_l(r_0) K_{a,b}^l(r) \; , \;\;
\Kcross(r_0,r)   =  \sum_{l=1}^{L} c_l(r_0) K_{b,a}^l(r) \, ,\\
\Kpavg(r_0,r)  &=& \sum_{l=1}^{L} c'_l(r_0) K_{b,a}^l(r) \; , \;\;
\Kpcross(r_0,r)  =  \sum_{l=1}^{L} c'_l(r_0) K_{a,b}^l(r) \, ,
\end{eqnarray}
which help to relate the inverted structural functions at $r_0$ to the true
structural functions:
\begin{eqnarray}
\left(\frac{\delta a}{a}\right)_{\mathrm{inv}}(r_0) &=& \int_0^R \left[\Kavg(r_0,r) \frac{\delta a(r)}{a(r)}
                               +  \Kcross(r_0,r) \frac{\delta b(r)}{b(r)} \right] \mathrm{d}r \, , \\
\left(\frac{\delta b}{b}\right)_{\mathrm{inv}}(r_0) &=& \int_0^R \left[\Kpcross(r_0,r) \frac{\delta a(r)}{a(r)}
                               +  \Kpavg(r_0,r) \frac{\delta b(r)}{b(r)} \right] \mathrm{d}r \, ,
\end{eqnarray}
where we have neglected the contribution from surface effects and observational
errors. As can be seen from these equations, the cross-term kernels help to
quantify the amount of cross-talk between the two functions in the inversion.

In the particular case of solar inversions, where the mass is known through
independent considerations, it is possible to constrain the inversion to
preserve the mass by introducing a supplementary Lagrange multiplier, provided
one of the structural variables being inverted is the density variation, $\delta
\rho_0/\rho_0$.  Indeed, if the mass is constant, then $\delta \rho_0/\rho_0$
obeys the following relation:
\begin{equation}
0 = 4 \pi \int_0^R \rho_0(r) \frac{\delta\rho_0}{\rho_0} r^2 \mathrm{d}r \, .
\label{eq:constant_mass}
\end{equation}

\subsubsection{Optimally Localised Averages (OLA)}

The OLA methods will also be modified due to the presence of two functions which
are being inverted.  Given that the modifications to the MOLA and SOLA variants
are similar, we will focus on the SOLA method in what follows.  First of all,
there will be two separate inversions, one for each of the functions being
inverted.  Secondly, not only do the averaging kernels need to be optimised, but
the cross-term kernels need to be reduced as much as possible.  These
considerations lead to cost functions of the following form:
\begin{eqnarray}
J(c_l(r_0)) &=& \int_0^R \left\{ \mathcal{T}(r_0,r) - \Kavg(r_0,r) \right\}^2 \mathrm{d}r +
\beta \int_0^R \left\{ \Kcross(r_0,r) \right\}^2 \mathrm{d}r  \nonumber \\
&+& \frac{\tan\theta \sum_{l=1}^L \left(c_l(r_0) \sigma_l\right)^2}{\left< \sigma^2 \right>}
 +  \lambda \left\{1- \int_0^R \Kavg(r_0,r)\mathrm{d}r \right\} \, .
\end{eqnarray}
For each inversion, there is a regularisation parameter ($\theta$), a
supplementary parameter to adjust the trade-off between optimising the averaging
kernel or minimising the cross-term kernel ($\beta$), a Lagrange multiplier to
ensure the averaging kernel is unimodular ($\lambda$), and optionally some
supplementary Lagrange multipliers used to suppress surface effects
\citep{Daeppen1991}.  The target functions ($\mathcal{T}$) for each of the
inverted functions can be adjusted independently.

In order to preserve the mass, for instance in the case of solar inversions,
one can treat Eq.~(\ref{eq:constant_mass}) as a supplementary observed
relation.  Specifically, $0$ will play the role of $\delta \omega/\omega$
and the function $f(r)=4\pi \rho r^2$ will be the kernel associated
with the structural variable $\delta \rho_0/\rho_0$.

\subsubsection{Applications}

\begin{figure}[t]
\centering
\includegraphics[width=0.45\textwidth]{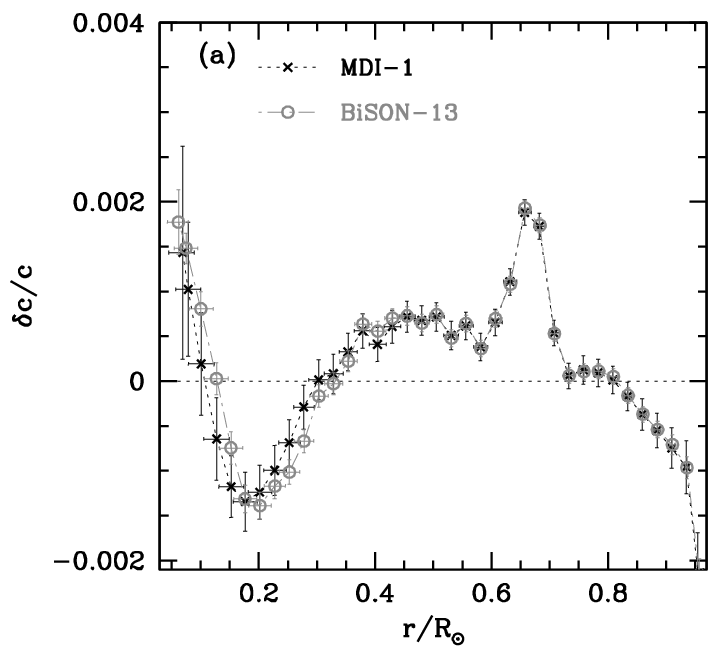} 
\includegraphics[width=0.45\textwidth]{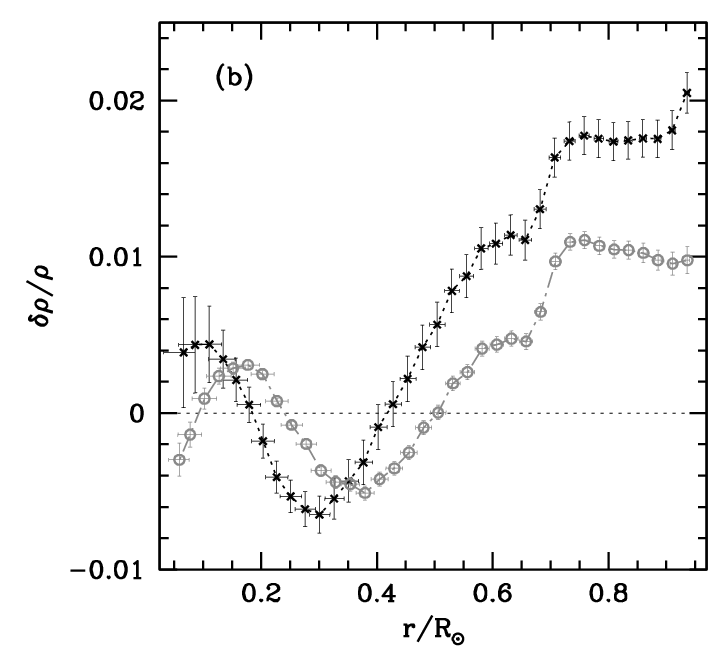}
\caption{ $(c,\rho)$ structural inversion for the Sun based on the results of
\citet{Basu2009}. Figure courtesy of S.~Basu. \label{fig:structural_inversion}}
\end{figure}

Up to now, structural inversions have been applied primarily to the Sun.
Figure~\ref{fig:structural_inversion}, which is based on the results of
\citet{Basu2009}, shows an example of such an inversion for the structural
variables $(c,\rho)$.  In recent years, the downward revision of the solar
metal abundances \citep[e.g.,][]{Asplund2009} has led to a significant
discrepancy between the results from solar structural inversions and models
based on these new abundances \citep[e.g.,][]{Basu2015}.  Indeed, helioseismic
inversions led to a lower depth for the base of the convection zone compared to
what is obtained from models with the revised abundances.  Currently, it is not
entirely clear how to solve this problem but different solutions are being
investigated.

\subsection{Integrated quantities}

\begin{figure}[t]
\centerline{\includegraphics[width=1.0\textwidth,clip,trim={35 440 0 0}]{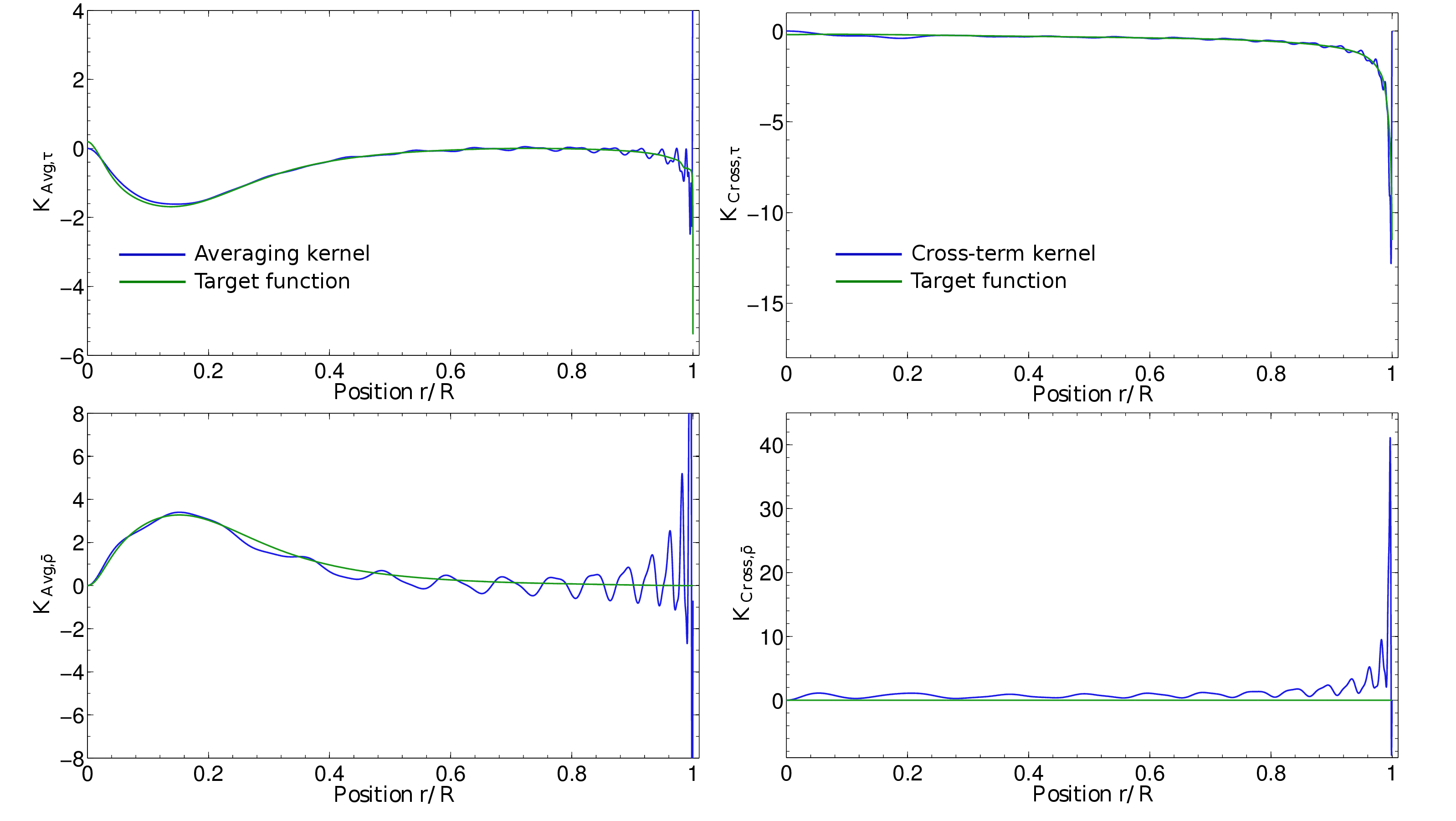}}
\caption{Acoustic radius inversion based on \citet{Buldgen2015a}.
Figure courtesy of G.~Buldgen. \label{fig:acoustic_radius_inversion}}
\end{figure}

In the case of stars other than the Sun, it is very difficult to carry out
structural inversions due to the limited number of available modes
\citep[e.g.,][]{Basu2002}.  Indeed, because of cancellation effects in
disc-integrated observations, only modes for which $\l \leq 3$ are detected
\citep{Dziembowski1977}. One strategy in such a situation is to invert stellar
parameters rather than structural profiles.   Indeed, since structural
inversions at a given grid point actually give a weighted average of the true
underlying profile, one can use a SOLA inversion to directly target the appropriate weight function
which yields the desired stellar parameter. The quantities which may be inverted
by such a procedure include the total angular momentum \citep{Pijpers1998}, the
mean density \citep{Reese2012}, the acoustic radius and various core or internal
mixing indicators \citep{Buldgen2015a, Buldgen2015b, Buldgen2017}.
Figure~\ref{fig:acoustic_radius_inversion} shows an inversion of the acoustic
radius, as described in \citet{Buldgen2015a}.

\section{Conclusion}

As illustrated in this course, inversions can be used to probe stellar rotation
profiles, probe the internal structure of stars, estimate various stellar
parameters, and indirectly test new physics outside a given grid of stellar
models.  Nonetheless, one must also bear the limitations of seismic inversions,
namely, the use of a priori assumptions about the smoothness of rotation or
structural profiles and the linearisation of the relationship between
frequencies and stellar structure (except in the case of non-linear inversions).
Furthermore, it is important to keep in mind that inversions cannot yield more
information than what is intrinsically contained in the observed pulsation
modes.

In order to get a more in-depth understanding of inversions, we recommend the
following articles or publications:
\begin{itemize}
\item \citet{Lynden-Bell1967} and \citet{Christensen-Dalsgaard2003}: the variational principle
\item \citet{Gough1991}: structural kernels
\item \citet{Christensen-Dalsgaard1990}: error propagation and magnification,
      averaging kernels
\item \citet{Rabello-Soares1999}: adjusting the free parameters in inversions
\item \citet{Reese2012} and \citet{Buldgen2015a}: inversions of integrated
      quantities
\end{itemize}
We also note that the recent monograph by \citet{Pijpers2006} contains several
chapters on helioseismic and asteroseismic inversions.

Various seismic inversion software packages have also become freely available in recent years:
\begin{itemize}
  \item \textsc{InversionKit}\footnote{See \url{http://bison.ph.bham.ac.uk/spaceinn/inversionkit}.}:
        1D inversions on individual stars
  \item \textsc{InversionPipeline}\footnote{See \url{http://bison.ph.bham.ac.uk/spaceinn/inversionpipeline}.}:
        inversions of stellar parameters using a grid of models
  \item \textsc{NonLinearKit}\footnote{See \url{http://bison.ph.bham.ac.uk/spaceinn/nonlinearkit}.}:
        non-linear 1D inversion tool still under development
  \item \textsc{SOLA Pack}\footnote{See \url{http://sun.stanford.edu/~rmunk/SOLApack/index.html}.}:
        2D rotation inversions in the Sun
\end{itemize}

\begin{acknowledgement}
I would like to thank the organisers of this Summer School for giving me the opportunity
to give this lecture on inversions.  Furthermore, I thank M.~J.~Thompson for
introducing me to inversions, as well as S.~Basu and G.~Buldgen for many
discussions on the topic.
\end{acknowledgement}

\bibliographystyle{apj}
\bibliography{biblio}

\begin{thebibliography}{}
\expandafter\ifx\csname natexlab\endcsname\relax\def\natexlab#1{#1}\fi

\bibitem[{{Asplund} {et~al.}(2009){Asplund}, {Grevesse}, {Sauval}, \&
  {Scott}}]{Asplund2009}
{Asplund}, M., {Grevesse}, N., {Sauval}, A.~J., \& {Scott}, P. 2009, ARA\&A,
  47, 481

\bibitem[{{Backus} \& {Gilbert}(1968)}]{Backus1968}
{Backus}, G., \& {Gilbert}, F. 1968, Geophysical Journal, 16, 169

\bibitem[{{Basu} {et~al.}(2009){Basu}, {Chaplin}, {Elsworth}, {New}, \&
  {Serenelli}}]{Basu2009}
{Basu}, S., {Chaplin}, W.~J., {Elsworth}, Y., {New}, R., \& {Serenelli}, A.~M.
  2009, ApJ, 699, 1403

\bibitem[{{Basu} \& {Christensen-Dalsgaard}(1997)}]{Basu1997}
{Basu}, S., \& {Christensen-Dalsgaard}, J. 1997, A\&A, 322, L5

\bibitem[{{Basu} {et~al.}(2002){Basu}, {Christensen-Dalsgaard}, \&
  {Thompson}}]{Basu2002}
{Basu}, S., {Christensen-Dalsgaard}, J., \& {Thompson}, M.~J. 2002, in ESA
  Special Publication, Vol. 485, Stellar Structure and Habitable Planet
  Finding, ed. B.~{Battrick}, F.~{Favata}, I.~W. {Roxburgh}, \& D.~{Galadi},
  249--252

\bibitem[{{Basu} {et~al.}(2015){Basu}, {Grevesse}, {Mathis}, \&
  {Turck-Chi{\`e}ze}}]{Basu2015}
{Basu}, S., {Grevesse}, N., {Mathis}, S., \& {Turck-Chi{\`e}ze}, S. 2015,
  {Space Science Reviews}, 196, 49

\bibitem[{{Bazot} {et~al.}(2012){Bazot}, {Bourguignon}, \&
  {Christensen-Dalsgaard}}]{Bazot2012}
{Bazot}, M., {Bourguignon}, S., \& {Christensen-Dalsgaard}, J. 2012, MNRAS,
  427, 1847

\bibitem[{{Brun} {et~al.}(2004){Brun}, {Miesch}, \& {Toomre}}]{Brun2004}
{Brun}, A.~S., {Miesch}, M.~S., \& {Toomre}, J. 2004, ApJ, 614, 1073

\bibitem[{{Buldgen} {et~al.}(2015{\natexlab{a}}){Buldgen}, {Reese}, \&
  {Dupret}}]{Buldgen2015b}
{Buldgen}, G., {Reese}, D.~R., \& {Dupret}, M.~A. 2015{\natexlab{a}}, A\&A,
  583, A62

\bibitem[{{Buldgen} {et~al.}(2017){Buldgen}, {Reese}, \&
  {Dupret}}]{Buldgen2017}
---. 2017, A\&A, 598, A21

\bibitem[{{Buldgen} {et~al.}(2015{\natexlab{b}}){Buldgen}, {Reese}, {Dupret},
  \& {Samadi}}]{Buldgen2015a}
{Buldgen}, G., {Reese}, D.~R., {Dupret}, M.~A., \& {Samadi}, R.
  2015{\natexlab{b}}, A\&A, 574, A42

\bibitem[{{Ceillier} {et~al.}(2013){Ceillier}, {Eggenberger}, {Garc{\'{\i}}a},
  \& {Mathis}}]{Ceillier2013}
{Ceillier}, T., {Eggenberger}, P., {Garc{\'{\i}}a}, R.~A., \& {Mathis}, S.
  2013, A\&A, 555, A54

\bibitem[{{Charpinet} {et~al.}(2005){Charpinet}, {Fontaine}, {Brassard},
  {Green}, \& {Chayer}}]{Charpinet2005}
{Charpinet}, S., {Fontaine}, G., {Brassard}, P., {Green}, E.~M., \& {Chayer},
  P. 2005, A\&A, 437, 575

\bibitem[{{Christensen-Dalsgaard}(2003)}]{Christensen-Dalsgaard2003}
{Christensen-Dalsgaard}, J. 2003, {Lecture Notes on Stellar Oscillations},
  \url{http://astro.phys.au.dk/~jcd/oscilnotes/}

\bibitem[{{Christensen-Dalsgaard}(2008)}]{Christensen-Dalsgaard2008}
---. 2008, ApSS, 316, 113

\bibitem[{{Christensen-Dalsgaard} {et~al.}(1990){Christensen-Dalsgaard},
  {Schou}, \& {Thompson}}]{Christensen-Dalsgaard1990}
{Christensen-Dalsgaard}, J., {Schou}, J., \& {Thompson}, M.~J. 1990, MNRAS,
  242, 353

\bibitem[{{D{\"a}ppen} {et~al.}(1991){D{\"a}ppen}, {Gough}, {Kosovichev}, \&
  {Thompson}}]{Daeppen1991}
{D{\"a}ppen}, W., {Gough}, D.~O., {Kosovichev}, A.~G., \& {Thompson}, M.~J.
  1991, in Lecture Notes in Physics, Berlin Springer Verlag, Vol. 388,
  Challenges to Theories of the Structure of Moderate-Mass Stars, ed.
  D.~{Gough} \& J.~{Toomre}, 111

\bibitem[{{Deheuvels} {et~al.}(2012){Deheuvels}, {Garc{\'{\i}}a}, {Chaplin},
  {Basu}, {Antia}, {Appourchaux}, {Benomar}, {Davies}, {Elsworth}, {Gizon},
  {Goupil}, {Reese}, {Regulo}, {Schou}, {Stahn}, {Casagrande},
  {Christensen-Dalsgaard}, {Fischer}, {Hekker}, {Kjeldsen}, {Mathur}, {Mosser},
  {Pinsonneault}, {Valenti}, {Christiansen}, {Kinemuchi}, \&
  {Mullally}}]{Deheuvels2012}
{Deheuvels}, S., {Garc{\'{\i}}a}, R.~A., {Chaplin}, W.~J., {et~al.} 2012, ApJ,
  756, 19

\bibitem[{{Deheuvels} {et~al.}(2014){Deheuvels}, {Do{\u g}an}, {Goupil},
  {Appourchaux}, {Benomar}, {Bruntt}, {Campante}, {Casagrande}, {Ceillier},
  {Davies}, {De Cat}, {Fu}, {Garc{\'{\i}}a}, {Lobel}, {Mosser}, {Reese},
  {Regulo}, {Schou}, {Stahn}, {Thygesen}, {Yang}, {Chaplin},
  {Christensen-Dalsgaard}, {Eggenberger}, {Gizon}, {Mathis},
  {Molenda-{\.Z}akowicz}, \& {Pinsonneault}}]{Deheuvels2014}
{Deheuvels}, S., {Do{\u g}an}, G., {Goupil}, M.~J., {et~al.} 2014, A\&A, 564,
  A27

\bibitem[{{Dziembowski}(1977)}]{Dziembowski1977}
{Dziembowski}, W. 1977, Acta Astronomica, 27, 203

\bibitem[{{Eggenberger} {et~al.}(2012){Eggenberger}, {Montalb{\'a}n}, \&
  {Miglio}}]{Eggenberger2012}
{Eggenberger}, P., {Montalb{\'a}n}, J., \& {Miglio}, A. 2012, A\&A, 544, L4

\bibitem[{{Elliott}(1996)}]{Elliott1996}
{Elliott}, J.~R. 1996, MNRAS, 280, 1244

\bibitem[{{Gough}(1985)}]{Gough1985}
{Gough}, D. 1985, Sol. Phys., 100, 65

\bibitem[{{Gough} \& {Thompson}(1991)}]{Gough1991}
{Gough}, D.~O., \& {Thompson}, M.~J. 1991, in Solar Interior and Atmosphere,
  ed. {Cox, A.~N., Livingston, W.~C., \& Matthews, M.~S.} (University of
  Arizona Press, Tucson), 519--561

\bibitem[{{Kosovichev}(1999)}]{Kosovichev1999}
{Kosovichev}, A.~G. 1999, Journal of Computational and Applied Mathematics,
  109, 1

\bibitem[{{Ledoux}(1951)}]{Ledoux1951}
{Ledoux}, P. 1951, ApJ, 114, 373

\bibitem[{{Lynden-Bell} \& {Ostriker}(1967)}]{Lynden-Bell1967}
{Lynden-Bell}, D., \& {Ostriker}, J.~P. 1967, MNRAS, 136, 293

\bibitem[{{Marques} {et~al.}(2013){Marques}, {Goupil}, {Lebreton}, {Talon},
  {Palacios}, {Belkacem}, {Ouazzani}, {Mosser}, {Moya}, {Morel}, {Pichon},
  {Mathis}, {Zahn}, {Turck-Chi{\`e}ze}, \& {Nghiem}}]{Marques2013}
{Marques}, J.~P., {Goupil}, M.~J., {Lebreton}, Y., {et~al.} 2013, A\&A, 549,
  A74

\bibitem[{{Masters}(1979)}]{Masters1979}
{Masters}, G. 1979, Geophysical Journal, 57, 507

\bibitem[{{Metcalfe} \& {Charbonneau}(2003)}]{Metcalfe2003}
{Metcalfe}, T.~S., \& {Charbonneau}, P. 2003, Journal of Computational Physics,
  185, 176

\bibitem[{{Pijpers}(1998)}]{Pijpers1998}
{Pijpers}, F.~P. 1998, MNRAS, 297, L76

\bibitem[{{Pijpers}(2006)}]{Pijpers2006}
---. 2006, {Methods in helio- and asteroseismology} (Imperial College Press)

\bibitem[{{Pijpers} \& {Thompson}(1992)}]{Pijpers1992}
{Pijpers}, F.~P., \& {Thompson}, M.~J. 1992, A\&A, 262, L33

\bibitem[{{Rabello-Soares} {et~al.}(1999){Rabello-Soares}, {Basu}, \&
  {Christensen-Dalsgaard}}]{Rabello-Soares1999}
{Rabello-Soares}, M.~C., {Basu}, S., \& {Christensen-Dalsgaard}, J. 1999,
  MNRAS, 309, 35

\bibitem[{{Reese}(2006)}]{Reese2006phd}
{Reese}, D. 2006, PhD thesis, Universit{\'e} Toulouse III - Paul Sabatier,
  http://tel.archives-ouvertes.fr/tel-00120334

\bibitem[{{Reese} {et~al.}(2012){Reese}, {Marques}, {Goupil}, {Thompson}, \&
  {Deheuvels}}]{Reese2012}
{Reese}, D.~R., {Marques}, J.~P., {Goupil}, M.~J., {Thompson}, M.~J., \&
  {Deheuvels}, S. 2012, A\&A, 539, A63

\bibitem[{{Schou} {et~al.}(1994){Schou}, {Christensen-Dalsgaard}, \&
  {Thompson}}]{Schou1994}
{Schou}, J., {Christensen-Dalsgaard}, J., \& {Thompson}, M.~J. 1994, ApJ, 433,
  389

\bibitem[{{Schou} {et~al.}(1998){Schou}, {Antia}, {Basu}, {Bogart}, {Bush},
  {Chitre}, {Christensen-Dalsgaard}, {di Mauro}, {Dziembowski}, {Eff-Darwich},
  {Gough}, {Haber}, {Hoeksema}, {Howe}, {Korzennik}, {Kosovichev}, {Larsen},
  {Pijpers}, {Scherrer}, {Sekii}, {Tarbell}, {Title}, {Thompson}, \&
  {Toomre}}]{Schou1998}
{Schou}, J., {Antia}, H.~M., {Basu}, S., {et~al.} 1998, ApJ, 505, 390

\bibitem[{{Silva Aguirre} {et~al.}(2015){Silva Aguirre}, {Davies}, {Basu},
  {Christensen-Dalsgaard}, {Creevey}, {Metcalfe}, {Bedding}, {Casagrande},
  {Handberg}, {Lund}, {Nissen}, {Chaplin}, {Huber}, {Serenelli}, {Stello}, {Van
  Eylen}, {Campante}, {Elsworth}, {Gilliland}, {Hekker}, {Karoff}, {Kawaler},
  {Kjeldsen}, \& {Lundkvist}}]{SilvaAguirre2015}
{Silva Aguirre}, V., {Davies}, G.~R., {Basu}, S., {et~al.} 2015, MNRAS, 452,
  2127

\bibitem[{{Thompson}(1993)}]{Thompson1993}
{Thompson}, M.~J. 1993, in Astronomical Society of the Pacific Conference
  Series, Vol.~42, GONG 1992. Seismic Investigation of the Sun and Stars, ed.
  T.~M. {Brown}, 141

\bibitem[{{Thompson} {et~al.}(2003){Thompson}, {Christensen-Dalsgaard},
  {Miesch}, \& {Toomre}}]{Thompson2003}
{Thompson}, M.~J., {Christensen-Dalsgaard}, J., {Miesch}, M.~S., \& {Toomre},
  J. 2003, ARA\&A, 41, 599

\bibitem[{{Unno} {et~al.}(1989){Unno}, {Osaki}, {Ando}, {Saio}, \&
  {Shibahashi}}]{Unno1989}
{Unno}, W., {Osaki}, Y., {Ando}, H., {Saio}, H., \& {Shibahashi}, H. 1989,
  {Nonradial oscillations of stars} (Tokyo: University of Tokyo Press)

\end{thebibliography}

\end{document}